\documentclass[prl,twocolumn,superscriptaddress,preprintnumbers,amsmath,amssymb]{revtex4}
\usepackage{graphicx}
\usepackage{amsmath}
\usepackage{latexsym}
\usepackage{dcolumn}
\usepackage{bm}
\usepackage{float}
\usepackage{graphicx,here}
\usepackage{color}
\usepackage{hyperref}
\usepackage{bookmark}
\usepackage{color}
\hypersetup{
citecolor=blue
}

\emergencystretch=\maxdimen
\begin{document}

\affiliation{Institute of Physics, Chinese Academy of Sciences, Beijing 100190, China}
\affiliation{Hiroshima Synchrotron Radiation Center, Hiroshima University, 2-313 Kagamiyama, Higashi-Hiroshima 739-0046, Japan}
\affiliation{Department of Applied Physics and Chemical Engineering, Tokyo University of Agriculture and Technology, Koganei, Tokyo 184-8588, Japan}
\affiliation{School of Physical Sciences, University of Chinese Academy of Sciences, Beijing 100049, China}
\affiliation{Songshan Lake Materials Laboratory, Dongguan, Guangdong 523808, China}
\affiliation{Center of Materials Science and Optoelectronics Engineering, University of Chinese Academy of Sciences, Beijing 100049, China}

\title{\bf Observation of topological flat bands in the kagome semiconductor Nb$_3$Cl$_8$}

\author{Zhenyu Sun}
\affiliation{Institute of Physics, Chinese Academy of Sciences, Beijing 100190, China}
\affiliation{School of Physical Sciences, University of Chinese Academy of Sciences, Beijing 100049, China}
\author{Hui Zhou}
\affiliation{Institute of Physics, Chinese Academy of Sciences, Beijing 100190, China}
\affiliation{School of Physical Sciences, University of Chinese Academy of Sciences, Beijing 100049, China}
\author{Cuixiang Wang}
\affiliation{Institute of Physics, Chinese Academy of Sciences, Beijing 100190, China}
\affiliation{School of Physical Sciences, University of Chinese Academy of Sciences, Beijing 100049, China}
\author{Shiv Kumar}
\affiliation{Hiroshima Synchrotron Radiation Center, Hiroshima University, 2-313 Kagamiyama, Higashi-Hiroshima 739-0046, Japan}
\author{Daiyu Geng}
\affiliation{Institute of Physics, Chinese Academy of Sciences, Beijing 100190, China}
\affiliation{School of Physical Sciences, University of Chinese Academy of Sciences, Beijing 100049, China}
\author{Shaosheng Yue}
\affiliation{Institute of Physics, Chinese Academy of Sciences, Beijing 100190, China}
\affiliation{School of Physical Sciences, University of Chinese Academy of Sciences, Beijing 100049, China}
\author{Xin Han}
\affiliation{Institute of Physics, Chinese Academy of Sciences, Beijing 100190, China}
\affiliation{School of Physical Sciences, University of Chinese Academy of Sciences, Beijing 100049, China}
\author{Yuya Haraguchi}
\affiliation{Department of Applied Physics and Chemical Engineering, Tokyo University of Agriculture and Technology, Koganei, Tokyo 184-8588, Japan}
\author{Kenya Shimada}
\affiliation{Hiroshima Synchrotron Radiation Center, Hiroshima University, 2-313 Kagamiyama, Higashi-Hiroshima 739-0046, Japan}
\author{Peng Cheng}
\affiliation{Institute of Physics, Chinese Academy of Sciences, Beijing 100190, China}
\affiliation{School of Physical Sciences, University of Chinese Academy of Sciences, Beijing 100049, China}
\author{Lan Chen}
\affiliation{Institute of Physics, Chinese Academy of Sciences, Beijing 100190, China}
\affiliation{School of Physical Sciences, University of Chinese Academy of Sciences, Beijing 100049, China}
\affiliation{Songshan Lake Materials Laboratory, Dongguan, Guangdong 523808, China}
\author{Youguo Shi\footnote[1]{ygshi@iphy.ac.cn}}
\affiliation{Institute of Physics, Chinese Academy of Sciences, Beijing 100190, China}
\affiliation{School of Physical Sciences, University of Chinese Academy of Sciences, Beijing 100049, China}
\affiliation{Songshan Lake Materials Laboratory, Dongguan, Guangdong 523808, China}
\affiliation{Center of Materials Science and Optoelectronics Engineering, University of Chinese Academy of Sciences, Beijing 100049, China}
\author{Kehui Wu\footnote[2]{khwu@iphy.ac.cn}}
\affiliation{Institute of Physics, Chinese Academy of Sciences, Beijing 100190, China}
\affiliation{School of Physical Sciences, University of Chinese Academy of Sciences, Beijing 100049, China}
\affiliation{Songshan Lake Materials Laboratory, Dongguan, Guangdong 523808, China}
\author{Sheng Meng\footnote[3]{smeng@iphy.ac.cn}}
\affiliation{Institute of Physics, Chinese Academy of Sciences, Beijing 100190, China}
\affiliation{School of Physical Sciences, University of Chinese Academy of Sciences, Beijing 100049, China}
\author{Baojie Feng\footnote[4]{bjfeng@iphy.ac.cn}}
\affiliation{Institute of Physics, Chinese Academy of Sciences, Beijing 100190, China}
\affiliation{School of Physical Sciences, University of Chinese Academy of Sciences, Beijing 100049, China}

\date{\today}

\clearpage

\begin{abstract}
\section{Abstract}
The destructive interference of wavefunctions in a kagome lattice can give rise to topological flat bands (TFBs) with a highly degenerate state of electrons. Recently, TFBs have been observed in several kagome metals, including Fe$_3$Sn$_2$, FeSn, CoSn, and YMn$_6$Sn$_6$. Nonetheless, kagome materials that are both exfoliable and semiconducting are lacking, which seriously hinders their device applications. Herein, we show that Nb$_3$Cl$_8$, which hosts a breathing kagome lattice, is gapped out because of the absence of inversion symmetry, while the TFBs survive because of the protection of the mirror reflection symmetry. By angle-resolved photoemission spectroscopy measurements and first-principles calculations, we directly observe the TFBs and a moderate band gap in Nb$_3$Cl$_8$. By mechanical exfoliation, we successfully obtain monolayer Nb$_3$Cl$_8$ which is stable under ambient conditions. In addition, our calculations show that monolayer Nb$_3$Cl$_8$ have a magnetic ground state, thus providing opportunities to study the interplay between geometry, topology, and magnetism.
\\ \hspace*{\fill} \\
{\bf Keywords:} breathing kagome lattice, topological flat bands, semiconductor, ARPES, DFT calculations, mechanical exfoliation
\end{abstract}

\maketitle

The interplay between geometry, topology, and magnetism at the quantum level can give rise to rich physical properties, and the exploration of novel physics in nontrivial lattices is at the forefront of condensed matter physics \cite{LinZ2018,KangM2020NM,KangM2020NC,LiuZ2020,LiM2021,WeeksC2010,SunK2011,SlotMR2017,Kempkes2019,YinJ2020,LiuH2021}. A prototypical example is the kagome lattice that is composed of corner-sharing triangles [Fig. 1(a)]. Such a simple lattice has been intensively studied because of the emergence of topological band structures and frustration-driven spin-liquid states \cite{GuoHM2009,BalentsL2010,TangE2011,HanT2012}. Based on a simple $s$-orbital tight-binding (TB) model with nearest-neighbour hopping, the electronic band structure of the kagome lattice can be described as a Dirac cone capped with a TFB [Fig. 1(b)]. The eigenfunctions at neighbouring corners have opposite phases, which results in a phase cancellation for hopping to neighbouring sites [black arrows in Fig. 1(a)]. Therefore, the electronic state is geometrically confined within a single hexagon; this real-space electronic localization leads to a TFB with quenched kinetic energy. The strong electron correlation effects in the highly degenerate flat band can give rise to various exotic properties, including high-temperature superconductivity \cite{ImadaM2000,JiangYX2021}, fractional quantum Hall effects \cite{TangE2011}, and Wigner crystal states \cite{JiangHC2017,JaworowskiB2018}.

Recently, TFBs and Dirac cones have been observed in several kagome metals by angle-resolved photoemission spectroscopy (ARPES), including Fe$_3$Sn$_2$ \cite{LinZ2018,YeL2018}, FeSn \cite{KangM2020NM}, CoSn \cite{KangM2020NC,LiuZ2020}, and YMn$_6$Sn$_6$ \cite{LiM2021}. However, most of the previously discovered kagome materials are metals without a band gap, which means that the ``OFF'' state cannot be achieved in devices. This drawback strongly limits their applications in logic and optoelectronic devices \cite{LiuY2016,LiuC2020}. Therefore, it is highly desirable to realize semiconducting kagome materials with a moderate band gap, in which the combination of semiconducting properties and TFBs might give rise to exotic physical phenomena, including gate-induced superconductivity \cite{YeJT2012} and triplet excitonic insulating state \cite{SethiG2021}. In addition, the realization of TFBs in semiconductors could enable the tuning of the TFBs by gating or chemical doping. On the other hand, an ideal kagome lattice is only one layer thick, and the non-negligible interlayer coupling in bulk materials will (partially) break the intrinsic properties of the kagome lattice. To date, experimental realization of layered and exfoliable kagome materials with TFBs is still challenging.

Here, we report a combined experimental and theoretical study on a breathing kagome material Nb$_3$Cl$_8$. Recently, the existence of TFBs in Nb$_3$Cl$_8$ has been predicted by a high-throughput screening \cite{LiuH2021}, but experimental investigations of its electronic structure are still lacking. Our ARPES and optical absorption spectroscopy measurements confirmed the existence of TFBs and a moderate band gap ($\sim$1.12 eV). By mechanical exfoliation, we successfully obtained monolayer Nb$_3$Cl$_8$, a key step for device applications. In addition, Nb$_3$Cl$_8$ is expected to have a magnetic ground state, which implies the existence of exotic properties arising from the interplay between magnetism and band topology.

\begin{figure*}[tbh]
\centering
\includegraphics[width=16 cm]{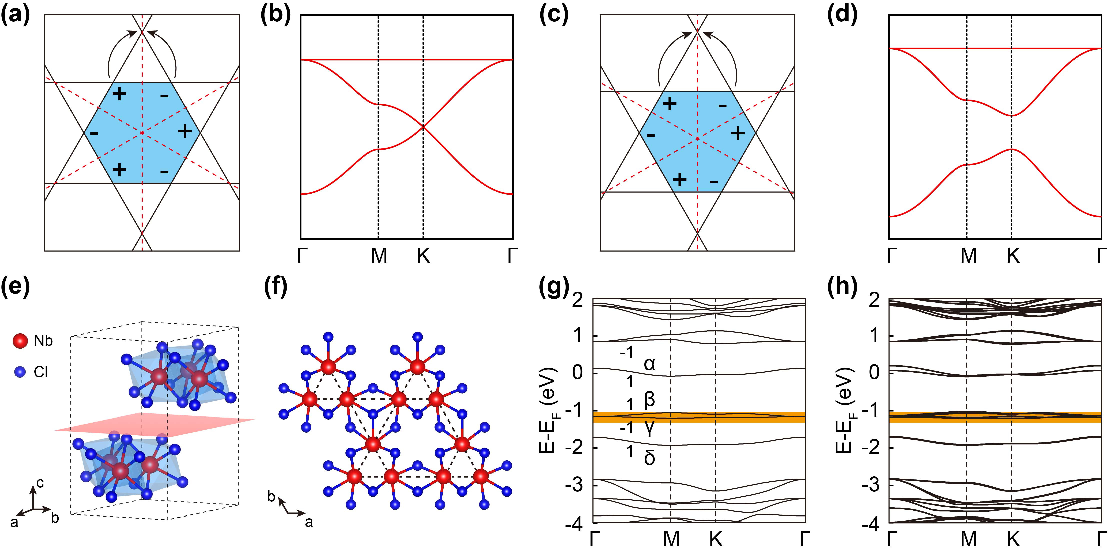}
\caption{{\bf Crystal and electronic structure of the kagome and breathing kagome lattices.} (a, c) Schematic drawing of the kagome lattice and breathing kagome lattice, respectively. Electrons are confined in the blue shaded hexagons because of destructive interference, as indicated by the black arrows. Red dashed lines indicate the mirror axis. (b, d) TB band structure of the kagome lattice and breathing kagome lattice, respectively. The (breathing) kagome lattice has a (gapped) Dirac cone and a topological flat band. (e) Three-dimensional crystal structure of Nb$_3$Cl$_8$. Red and blue balls indicate Nb and Cl atoms, respectively. (f) The crystal structure of monolayer Nb$_3$Cl$_8$. The Nb atoms form a breathing kagome lattice, as indicated by the black dashed lines. (g, h) The calculated band structures of monolayer (g) and bulk (h) Nb$_3$Cl$_8$ in the paramagnetic state. The parity of the mirror operator along $\Gamma$-M is labelled by ``+'' and ``-'' near each band. The four bands that have been observed by ARPES measurements are indicated by $\alpha$, $\beta$, $\gamma$, and $\delta$, respectively.}
\end{figure*}

First, we briefly discuss the band structure of the breathing kagome lattice [Fig. 1(c)] based on a TB model. The degeneracy of the Dirac point in the conventional kagome lattice is protected by inversion symmetry. In the breathing kagome lattice, however, the absence of inversion symmetry will gap out the Dirac cone, leading to a semiconducting ground state, as shown in Fig. 1(d). On the other hand, destructive phase interference, the key reason for the emergence of TFBs, is protected by three equivalent mirror axes (red dashed lines in Fig. 1(a) and 1(c). These mirror symmetries survive in the breathing kagome lattice, and thus, the breathing kagome lattice still hosts TFBs. A detailed TB analysis is presented in the Supplementary Materials.

\begin{figure*}[htb]
\centering
\includegraphics[width=16 cm]{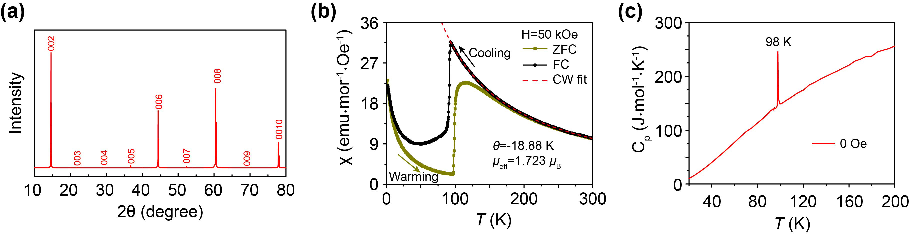}
\caption{{\bf Characterization of the basic physical properties of Nb$_3$Cl$_8$.} (a) X-ray diffraction pattern of the Nb$_3$Cl$_8$ single crystal. (b) Temperature dependence of the magnetic susceptibility ($\chi$) of Nb$_3$Cl$_8$. The red dashed line indicates the fitting results using the Curie-Weiss function. (c) Temperature dependence of the heat capacity of Nb$_3$Cl$_8$.}
\end{figure*}

Next, we study a van der Waals material Nb$_3$Cl$_8$, which has a breathing kagome lattice in each layer \cite{MillerGJ1995,HaraguchiY2017,YoonJ2020,PascoCM2020,SheckeltonJP2017}. Bulk Nb$_3$Cl$_8$ crystallizes in a layered structure with a space group of {\it P\={3}m1} [Fig. 1(e)]. Within each layer, the Nb atoms form a breathing kagome lattice, as shown in Fig. 1(f). The calculated band structure of monolayer Nb$_3$Cl$_8$ is displayed in Fig. 1(g). One can directly see a flat band at approximately 1.1 eV below the Fermi level, which has the narrowest bandwidth, as highlighted by the orange region. In addition, our orbital and symmetry analysis has confirmed the phase destructive interfering behavior of this band, which is the key character of kagome TFBs (see Supporting Information for details). The Dirac cones are gapped out, resulting in multiple gaps. Notably, the $\alpha$ band crosses the Fermi level, indicating a metallic ground state. However, we will show later that the ground state should be semiconducting because DFT calculations underestimate the gap, as confirmed by our ARPES and optical absorption spectroscopy measurements. Because of the weak interlayer van der Waals interaction, the band structure of bulk Nb$_3$Cl$_8$ is analogous to that of monolayer Nb$_3$Cl$_8$, as shown in Fig. 1(h). As a result, the TFBs survive in bulk Nb$_3$Cl$_8$, which is favourable for ARPES measurements.

High-quality Nb$_3$Cl$_8$ single crystals were synthesized by the self-flux method (see the Methods section). Figure 2(a) shows an XRD spectrum measured along the (\emph{00l}) direction, and only the \emph{00l} peaks were observed. The sharp peaks indicate the high quality of the crystals. The temperature-dependent magnetic susceptibility of Nb$_3$Cl$_8$ is displayed in Fig. 2(b). As the temperature decreases, the magnetic susceptibility of Nb$_3$Cl$_8$ drops abruptly at $\sim$100 K with a prominent hysteresis. Above 100 K, the temperature dependence of the magnetic susceptibility can be well fitted by the Curie-Weiss function, as indicated by the red dashed line in Fig. 2(b), and the fitted $\theta$ value is -18.88 K. The negative $\theta$ value indicates an antiferromagnetic ground state. These results agree well with previous works \cite{HaraguchiY2017}. In addition, the heat capacity shows a sharp $\lambda$-shaped peak at approximately $\sim$98 K [Fig. 2(c)], in line with the transition temperature determined from the magnetic susceptibility measurements. This phase transition were interpreted as a slight change in layer stacking, which does not affect the structure within each layer \cite{HaraguchiY2017}. Therefore, the topological properties of Nb$_3$Cl$_8$ are not affected by the phase transition either.

We then performed ARPES measurements to confirm the TFBs in Nb$_3$Cl$_8$. Because of the semiconducting nature of Nb$_3$Cl$_8$, there is no detectable photoemission signal at the Fermi level. With increasing binding energies, we observe strong spectral weight near the $\Gamma$ point of the first Brillouin zone (BZ) from $E\rm_B\sim$0.8 eV, as shown in Fig. 3(a). The spectral weight in the second BZ is much weaker because of the photoemission matrix element effect. The constant energy contour at $E\rm_B\sim$2.6 eV shows clear hexagonal symmetry [Fig. 3(b)], in agreement with the crystal structure of Nb$_3$Cl$_8$. The band structures along the $\bar{\Gamma}$-$\bar{K}$ and $\bar{\Gamma}$-$\bar{M}$ directions are displayed in Fig. 3(c)-3(i). A careful comparison with the calculation results shows that the chemical potential has a $\sim$0.8 eV shift towards higher binding energies. After readjusting the chemical potential, the calculated band structures agree well with our ARPES results except for a slight discrepancy at the Fermi level, as indicated by the red dashed lines. The discrepancy at the Fermi level will be discussed later. Within 3.5 eV of the Fermi level, we observed four prominent bands: $\alpha$, $\beta$, $\gamma$, and $\delta$. Notably, the TFB, {\it i.e.}, the $\gamma$ band can be observed in the whole BZ and has negligible dispersion, as shown in Fig. 3(e),(f),(j). The $\beta$ band, which is very close to the $\gamma$ band, has a stronger spectral weight with low photon energies [Fig. 3(i)]. In contrast with the flat band, the $\beta$ band disperses with momentum, as shown in Fig. 3(k). Figure 3(l) shows the fitted dispersions of the four bands, which agree well with our calculation results.

\begin{figure*}[htb]
\centering
\includegraphics[width=16 cm]{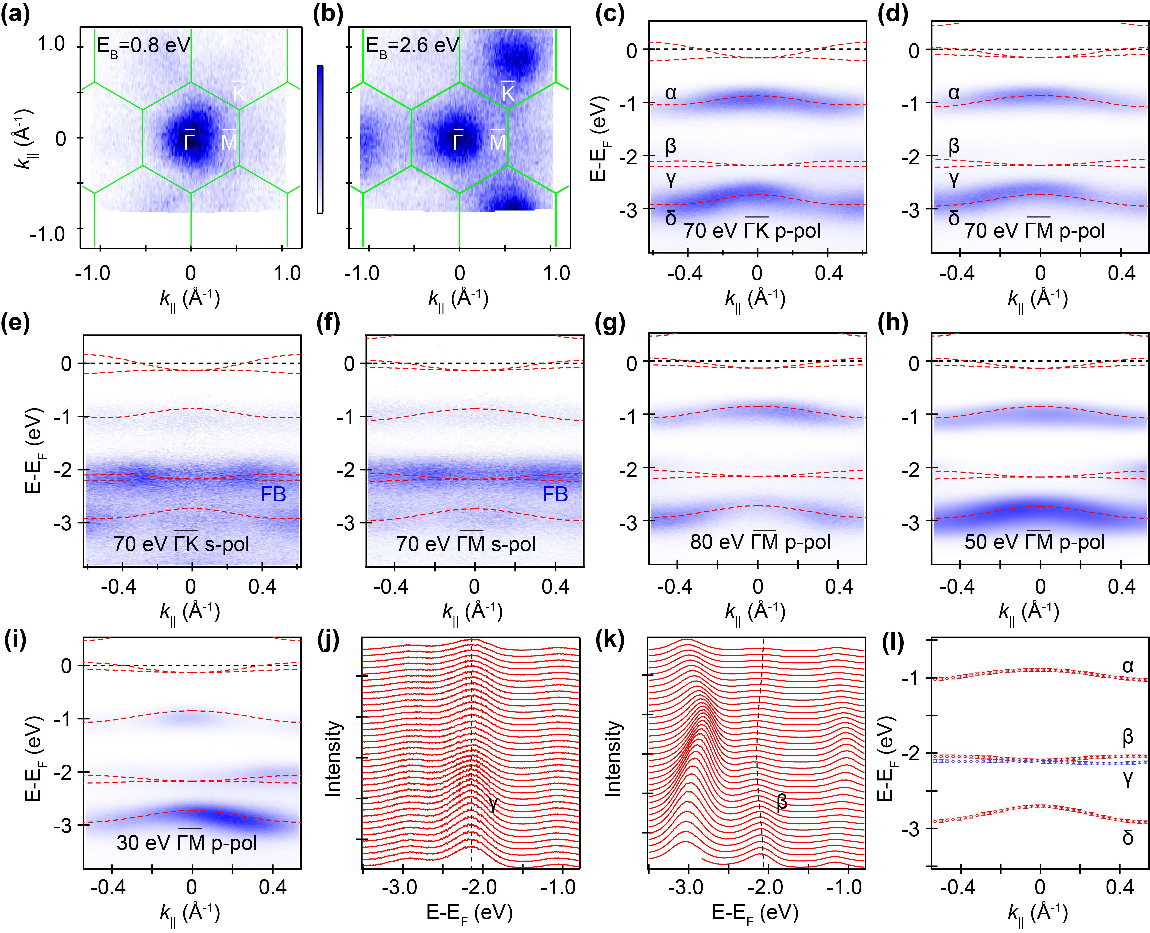}
\caption{{\bf ARPES measurements of Nb$_3$Cl$_8$.} (a,b) Constant energy contours at $E\rm_B$=0.8 and 2.6 eV, respectively. Green lines indicate the BZs of monolayer Nb$_3$Cl$_8$. (c-i) ARPES intensity plots along the $\bar{\Gamma}$-$\bar{K}$ and $\bar{\Gamma}$-$\bar{M}$ directions with different photon energies and polarizations. The calculated band structures of monolayer Nb$_3$Cl$_8$ are superimposed on each panel after shifting the Fermi level 0.8 eV towards a higher binding energy. (j,k) Energy distribution curves of (f) and (i), respectively. Black dashed lines indicate the $\gamma$ and $\beta$ bands, respectively. (l) Fitted dispersions of the four bands based on the peaks in energy distribution curves. The $\alpha$ and $\delta$ bands are fitted using the data in (c); the $\gamma$ band is fitted using the data in (f); the $\beta$ band is fitted using the data in (h).}
\end{figure*}

During ARPES measurements, we find that the spectral weight of each band varies with the polarization of the incident light: the $\alpha$, $\beta$, and $\delta$ bands are more sensitive to $p$-polarized light, while the $\gamma$ band is more sensitive to $s$-polarized light, despite the intensity variation with photon energy. To understand this phenomenon, we focus on the bands along $\bar{\Gamma}$-$\bar{M}$ because $\bar{\Gamma}$-$\bar{M}$ is a mirror axis and each band along this direction has definite odd or even parity when spin-orbit coupling is neglected. The calculation results are indicated by ``+'' (even) and ``-"  (odd) in Fig. 1(g). We find that the $\alpha$, $\beta$, and $\delta$ bands have even parity, which can only be observed by $p$-polarized light. The $\gamma$ band, {\it i.e.}, the flat band has odd parity and can only be detected by $s$-polarized light. These analyses agree well with our ARPES results, which further confirms the existence of TFBs in Nb$_3$Cl$_8$. Neglecting the variation in relative intensity, we find that all bands have negligible dispersion with photon energy (or $k_z$), in agreement with the two-dimensional nature of Nb$_3$Cl$_8$.

Next, we discuss the discrepancy near the Fermi level between the experimental and calculation results: the calculation results show the existence of bands after adjusting the chemical potential, while we did not observe any bands in our ARPES measurements. A possible reason for this discrepancy is the underestimation of the band gap in DFT calculations \cite{PerdewJP2017}. Based on our ARPES results, the valence band maximum is located at $\sim$0.9 eV below the Fermi level, which indicates that the actual band gap is larger than 0.9 eV. To determine the band gap of Nb$_3$Cl$_8$, we performed optical absorption spectroscopy measurements. Figure 4(a) shows a typical absorption spectrum of Nb$_3$Cl$_8$, and several absorption peaks are observed from the ultraviolet to near-infrared regions. There is one prominent peak at 1.12 eV, which can be assigned to the transition from valence to conduction bands, as indicated by the blue arrow in Fig. 4(b). Therefore, the optical band gap is $\sim$1.12 eV, which is approximately 0.23 eV larger than the calculation results. It should be noted that there is a broader peak centred at $\sim$6250 cm$^{-1}$ (0.775 eV). This feature is within the band gap and might originate from defects or edge states \cite{RoxloCB1986,ChenJ2021}. Since the valence band maximum is located at $E\rm_B\sim$0.9 eV, the conduction band bottom is expected to be located at $>$0.22 eV above the Fermi level, which is not accessible in our ARPES measurements.

Thus far, we have demonstrated that Nb$_3$Cl$_8$ hosts TFBs because of the breathing kagome lattice. Strictly speaking, however, a kagome lattice refers to monolayer materials; the non-negligible interlayer coupling in bulk materials could be detrimental to the intrinsic properties of the kagome lattice. Therefore, it is highly desirable to obtain monolayer kagome materials for their applications in quantum devices. Because of the weak van der Waals interaction in bulk Nb$_3$Cl$_8$, ultrathin flakes can be easily obtained by mechanical exfoliation \cite{YoonJ2020}. Figure 4(c) shows a typical optical microscope image of an ultrathin Nb$_3$Cl$_8$ flake, and Figure 4(d) shows an atomic force microscopy (AFM) image in the white box of Fig. 4(c). The height of the thinnest flake is approximately 0.8 nm with respect to the substrate [Fig. 4(e)], which is comparable to the lattice constant of bulk Nb$_3$Cl$_8$ in the perpendicular direction ($\sim$0.61 nm). The quality of ultrathin Nb$_3$Cl$_8$ was also confirmed by Raman spectroscopy measurements, as shown in Fig. 4(f). Most of the Raman peaks of the bulk Nb$_3$Cl$_8$ survive in the monolayer limit with negligible shift. It should be noted that our AFM and Raman spectroscopy measurements were performed under ambient conditions, and the strong intensity of Raman peaks indicates the high stability of monolayer Nb$_3$Cl$_8$ in air. Stability in air is an essential prerequisite for potential device applications.

\begin{figure}[htb]
\centering
\includegraphics[width=8 cm]{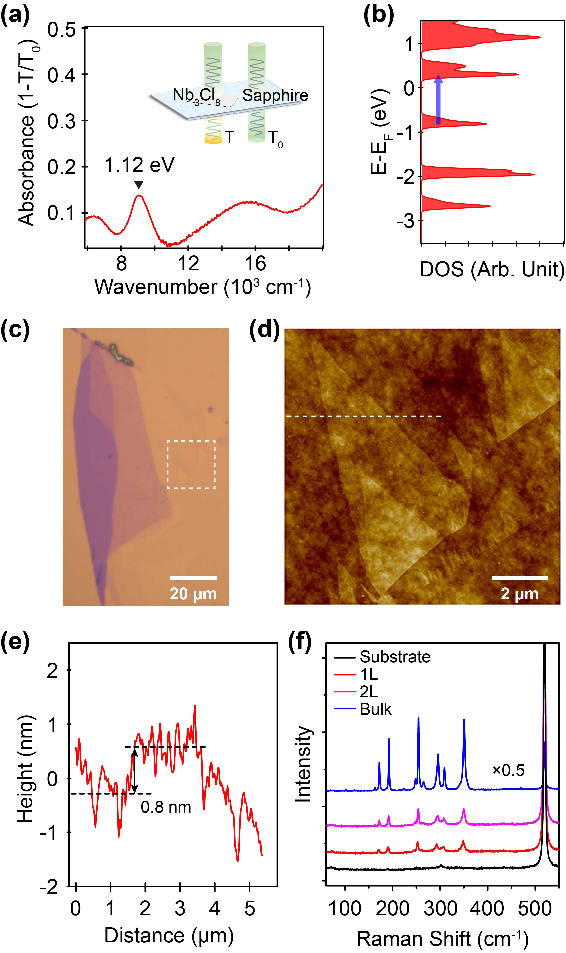}
\caption{{\bf Optical absorption measurements and mechanical exfoliation of Nb$_3$Cl$_8$.} (a) Extinction spectrum (1-T/T$_0$) of a 100-nm thick Nb$_3$Cl$_8$ flake. The supporting substrate is sapphire. (b) Calculated local density of states. The chemical potential and band gap were adjusted according to our experimental results. The blue arrow indicates the optical transition from the valence band maximum to the conduction band minimum. (c) Optical microscope image of an exfoliated Nb$_3$Cl$_8$ flake. (d) AFM image in the white box in (c). (e) Line profile along the white dashed line in (d). (f) Raman spectra of 1L, 2L, and bulk Nb$_3$Cl$_8$.}
\end{figure}

To summarize, we provide compelling evidence for the existence of TFBs in Nb$_3$Cl$_8$ which has a breathing kagome lattice. The semiconducting nature of Nb$_3$Cl$_8$ provides opportunities to fabricate optoelectronic and logic devices. Few- and monolayer Nb$_3$Cl$_8$ can be easily obtained by mechanical exfoliation and are quite stable under ambient conditions, which is an essential prerequisite for their device applications. In addition, our calculations show that the ground state of monolayer Nb$_3$Cl$_8$ is ferromagnetic and the TFBs will spin split because of the magnetic exchange interaction (see Supplementary Materials). In contrast with conventional and nonmagnetic kagome lattices, the breaking of symmetries in breathing kagome lattices, including inversion and time reversal, can give rise to multiple topological states, such as higher-order topological insulators/semimetals \cite{BolensA2019,WakaoH2020,EzawaM2018}, Chern insulators \cite{RenY2021}, and chiral charge densities \cite{JiangYX2021}.

\section{Methods}

Single crystals of Nb$_3$Cl$_8$ were grown using PbCl2 as a flux. High-purity Nb (Alfa Aesar 99.99\%) and NbCl$_5$ (Alfa Aesar 99.9\%) were mixed at a molar ratio of 7:8 and placed in an alumina crucible. The crucible was sealed in a quartz tube under vacuum, heated at 750 $^{\circ}$C for 150 hours, and cooled to room temperature naturally. The excess PbCl$_2$ flux was removed by sonicating in hot water. ARPES measurements were performed at Beamline BL-1 of the Hiroshima Synchrotron Radiation Center \cite{IwasawaH2017}. Clean surfaces required for the ARPES measurements were obtained by cleaving the samples in situ in an ultrahigh vacuum chamber. Both the cleavage and the measurements were performed at room temperature to avoid charging effects. Nb$_3$Cl$_8$ thin flakes can be exfoliated on various substrates, including SiO$_2$/Si, Au/SiO$_2$/Si, and sapphire, using blue Nitto tapes. Freshly cleaved surfaces were attached to the precleaned substrates, followed by carefully peeling off the tape. Thin flakes that contain few- to monolayer Nb$_3$Cl$_8$ remained on the substrates. The thicknesses and morphologies of Nb$_3$Cl$_8$ flakes were examined by AFM (Oxford, Asylum Research Cypher S) in tapping mode. Raman spectra were collected using a confocal Raman system (Horiba LanRam HR Evolution) with 532-nm laser excitation. Optical absorption measurements were carried out using a commercial spectrometer (MStarter ABS, Metatest Corporation).

First-principles calculations based on density functional theory were performed with the Vienna ab initio simulation package \cite{KresseG1993,KresseG1996prb}. The projector-augmented wave pseudopotential \cite{BlochlPE1994} and Perdew-Burke-Ernzerhof exchange-correlation functional \cite{PerdewJP1996} were used. The energy cutoff of the plane-wave basis was set at 350 eV. For monolayer Nb$_3$Cl$_8$, the vacuum space was set to be larger than 15 \AA. The first BZ was sampled according to the $\Gamma$-centred scheme. For monolayer (bulk) Nb$_3$Cl$_8$, we used a $k$ mesh of 6$\times$6$\times$1 (6$\times$6$\times$6) for structural optimization and 12$\times$12$\times$1 (9$\times$9$\times$9) for the self-consistent calculations. The positions of the atoms were optimized until the convergence of the force on each atom was less than 0.01 eV/\AA. The convergence condition of the electronic self-consistent loop was 10$^{-5}$ eV.
~\\

\section{Associated Content}

{\bf\noindent Supporting Information}\\
Generalizing the destructive phase interference to breathing kagome lattice; Inversion symmetry breaking induced gap opening in the kagome lattice; Determining the magnetic ground state of monolayer Nb$_3$Cl$_8$; Topological properties of the flat bands in Nb$_3$Cl$_8$; Orbital and symmetry analysis of the flat band.
~\\

\section{Acknowledgement}

This work was supported by the Ministry of Science and Technology of China (Grant No. 2018YFE0202700), the National Natural Science Foundation of China (Grants No. 11974391, No. 11825405, No. 1192780039, and No. U2032204), the Beijing Natural Science Foundation (Grant No. Z180007), the International Partnership Program of Chinese Academy of Sciences (Grant No. 112111KYSB20200012), and the Strategic Priority Research Program of Chinese Academy of Sciences (Grants No. XDB33030100). ARPES measurements were performed under the Proposal No. 21AU002. We thank the N-BARD, Hiroshima University for supplying liquid He.
~\\

\section{Author Contributions}

\noindent B.F. conceived the research. C.W., X.H., Y.H., and Y.S. synthesized the crystals; Z.S., S.K., D.G., S.Y., and B.F. performed the ARPES measurements; H.Z. and S.M. performed theoretical calculations and analysis; all authors contributed to the discussion of the data and writing of the manuscript. Z.S, H.Z., C.W., and S.K. contributed equally to this work.

\section{Notes}

\noindent The authors declare no competing financial interests.
~\\

\end{document}